\documentclass[a4paper,twoside,12pt]{article}
\input{obsjkt.sty}

\newcommand{\Msun}{\ensuremath{\,{\rm M}_\odot}}                  
\newcommand{\Rsun}{\ensuremath{\,{\rm R}_\odot}}                  
\newcommand{\rhosun}{\ensuremath{\,\rho_\odot}}                   
\newcommand{\Teff}{\ensuremath{T_{\rm eff}}}                      
\newcommand{\degr}{\ensuremath{^\circ}}                           
\renewcommand{\kms}{\,km\,s$^{-1}$}                               
\newcommand{\etal}{\textit{et al.}}                               

\newcommand{\Msunnom}{\hbox{$\mathcal{M}^{\rm N}_\odot$}}
\newcommand{\Rsunnom}{\hbox{$\mathcal{R}^{\rm N}_\odot$}}
\newcommand{\Lsunnom}{\hbox{$\mathcal{L}^{\rm N}_\odot$}}

\OBSheader{Rediscussion of eclipsing binaries: V1022 Cas}{J.\ Southworth}{2021 Apr}

\begin{document} 

\OBStitle{Rediscussion of eclipsing binaries. Paper III. \\ The interferometric, spectroscopic and eclipsing binary V1022\,Cassiopeiae}

\OBSauth{John Southworth}

\OBSinstone{Astrophysics Group, Keele University, Staffordshire, ST5 5BG, UK}

\OBSabstract{V1022\,Cas has been known as a spectroscopic binary for a century. It was found to be eclipsing based on photometry from the {\it Hipparcos} satellite, and an astrometric orbit was recently obtained from near-infrared interferometry. We present the first high-precision measurement of the radii of the stars based on light curves obtained by the TESS satellite. Combined with published radial velocities from high-resolution spectra, we measure the masses of the stars to be $1.626 \pm 0.001$\Msun\ and $1.609 \pm 0.001$\Msun, and the radii to be $2.591 \pm 0.026$\Rsun\ and $2.472 \pm 0.027$\Rsun. The 12.16\,d orbit is eccentric and the stars rotate sub-synchronously, so the system is tidally unevolved. A good match to these masses and radii, and published temperatures of the stars, is found for several sets of theoretical stellar evolutionary models, for a solar metallicity and an age of approximately 2\,Gyr. Four separate distance determinations to the system are available, and are in good agreement. The distances are based on surface brightness calibrations, theoretical bolometric corrections, the {\it Gaia} parallax, and the angular size of the astrometric orbit. A detailed spectroscopic analysis of the system to measure chemical abundances and more precise temperatures would be helpful.}


\section*{Introduction}

The study of binary stars is one of the oldest areas of astronomy. It can be traced back as far as John Michell \cite{Michell1767rspt}, who in 1767 concluded that the celestial positions of the bright stars were statistically improbable unless they were `placed near together'. William Herschel \cite{Herschel1802rspt,Herschel1803rspt} confirmed this by detecting orbital motion in several double stars that were visually resolved on the sky. This class of binary system is now typically referred to as visual binaries (if they are visually separable), astrometric binaries (if an astrometric orbit has been obtained), or interferometric binaries (if the stars have been spatially resolved using interferometry), although these terms are often interchanged.

A second type of binary system is the eclipsing binaries (EBs). The correct explanation for the periodic dimmings was originally suggested by John Goodricke \cite{Goodricke1783}, but observations of these can be traced back to antiquity \cite{JetsuPorceddu15plos}. A third type is the spectroscopic binaries, whose explanation is comparatively more recent as it required the development of astronomical spectroscopy. Vogel \cite{Vogel1890pasp} proved that the star Algol was a close binary by measuring the change in radial velocity (RV) of the primary star over a primary eclipse.

The definitions of these three types of binary system are observational, and this same nature determines what physical properties are measurable \cite{Me20conf}. An astrometric orbit for stars of a known distance gives the semimajor axis and the sum of the masses. The RV measurements that comprise a spectroscopic orbit gives lower limits on the masses of the stars\footnote{For the purposes of this discussion it is assumed that RVs of both stars are available.}. Observations of eclipses give the fractional radii of the stars (their radii divided by the semimajor axis of the relative orbit).

When a binary system can be studied using two methods, it is possible to determine many more physical properties. Astrometry and RVs together give the individual masses of the stars plus a direct geometric measurement of the distance to the system \cite{Torres++97apj2,Torres++97apj}. Eclipses and RVs allow the masses and radii of the stars to be measured to high precision \cite{Andersen91aarv,Torres++10aarv}.

A small number of binary systems are both astrometric, spectroscopic and eclipsing. For these objects it is possible to measure their masses, radii and distance to high precision through geometrical arguments alone. This list includes $\beta$\,Aurigae \cite{Stebbins11apj,Hummel+95aj,Me++07aa}, $\beta$\,Persei \cite{Vogel1890pasp,Zavala+10apj,Kolbas+15mn}, TZ\,For \cite{Andersen+91aa,Gallenne+16aa} and six other EBs recently studied by Gallenne \etal\ \cite{Gallenne+19aa}.


\begin{table}[t]
\caption{\em Basic information on V1022\,Cas \label{tab:info}}
\centering
\begin{tabular}{lll}
{\em Property}                      & {\em Value}            & {\em Reference}                   \\[3pt]
Bright Star Catalogue               & HR 9059                & \cite{HoffleitJaschek91}          \\
Henry Draper designation            & HD 224355              & \cite{CannonPickering24anhar}     \\
\textit{Gaia} DR2 ID                & 1994714276926012416    & \cite{Gaia18aa}                   \\
\textit{Gaia} parallax              & $15.767 \pm 0.088$ mas & \cite{Gaia18aa}                   \\
$B$ magnitude                       & $6.033 \pm 0.014$      & \cite{Hog+00aa}                   \\
$V$ magnitude                       & $5.561 \pm 0.009$      & \cite{Hog+00aa}                   \\
$J$ magnitude                       & $4.618 \pm 0.035$      & \cite{Cutri+03book}               \\
$H$ magnitude                       & $4.429 \pm 0.176$      & \cite{Cutri+03book}               \\
$K_s$ magnitude                     & $4.381 \pm 0.021$      & \cite{Cutri+03book}               \\
Spectral type                       & F6\,V + F6\,V          & \cite{Cowley76pasp}               \\[10pt]
\end{tabular}
\end{table}

\section*{V1022\,Cassiopeiae}

The current series of papers was conceived with the primary aim of measuring precise masses and radii of EBs that could then be included in DEBCat\footnote{\texttt{https://www.astro.keele.ac.uk/jkt/debcat/}} (Detached Eclipsing Binary Catalogue), a compilation of EBs with masses and radiii measured to precisions of 2\% or better \cite{Me15debcat}. Previous papers in this series have studied the B-type system $\zeta$\,Phoenicis \cite{Me20obs} and the solar-type system KX\,Cancri \cite{Me21obs1}.

In the current work the aim is to analyse the binary system V1022\,Cassiopeiae, one of a small number of spectroscopic-astrometric-eclipsing binaries, in order to determine its masses and radii to sufficient precision for inclusion in DEBCat. V1022\,Cas is a bright ($V=5.6$) binary system with an eccentric orbit of period 12.16\,d. The two stars are very similar, slightly evolved, and have spectral types of F6\,V \cite{Cowley76pasp}.

V1022\,Cas (Table\,\ref{tab:info}) was found to be a double-lined spectroscopic binary by Plaskett \etal\ \cite{Plaskett+20pdao} and several spectroscopic orbits have been published \cite{Harper23pdao,Imbert77aas,Fekel++10aj}. The last of these, by Fekel \etal\ \cite{Fekel++10aj}, is of very high quality.

Otero \etal\ \cite{Otero06ibvs} found the system to be eclipsing using photometry from the \textit{Hipparcos} satellite \cite{Hip97,Perryman+97aa}, which included four points around times of primary eclipse that were significantly fainter than the rest. No secondary eclipse was observed, although there is a marginal detection based on two datapoints \cite{Fekel++10aj}. The New catalogue of Suspected Variable Stars \cite{KukarkinKholopov82book} listed it as a suspected EB with an eclipse depth of only 0.05\,mag \cite{Otero06ibvs}.

Lester \etal\ \cite{Lester+19aj} have recently presented $K$-band interferometry of V1022\,Cas using the CHARA array, from which they obtained the first astrometric orbit of the system. They spatially resolved the two stars on nine occasions and, together with new and published RVs, determined the masses of the stars and the distance to the system to high precision.


\section*{Observational material}

\begin{figure}[t] \centering \includegraphics[width=\textwidth]{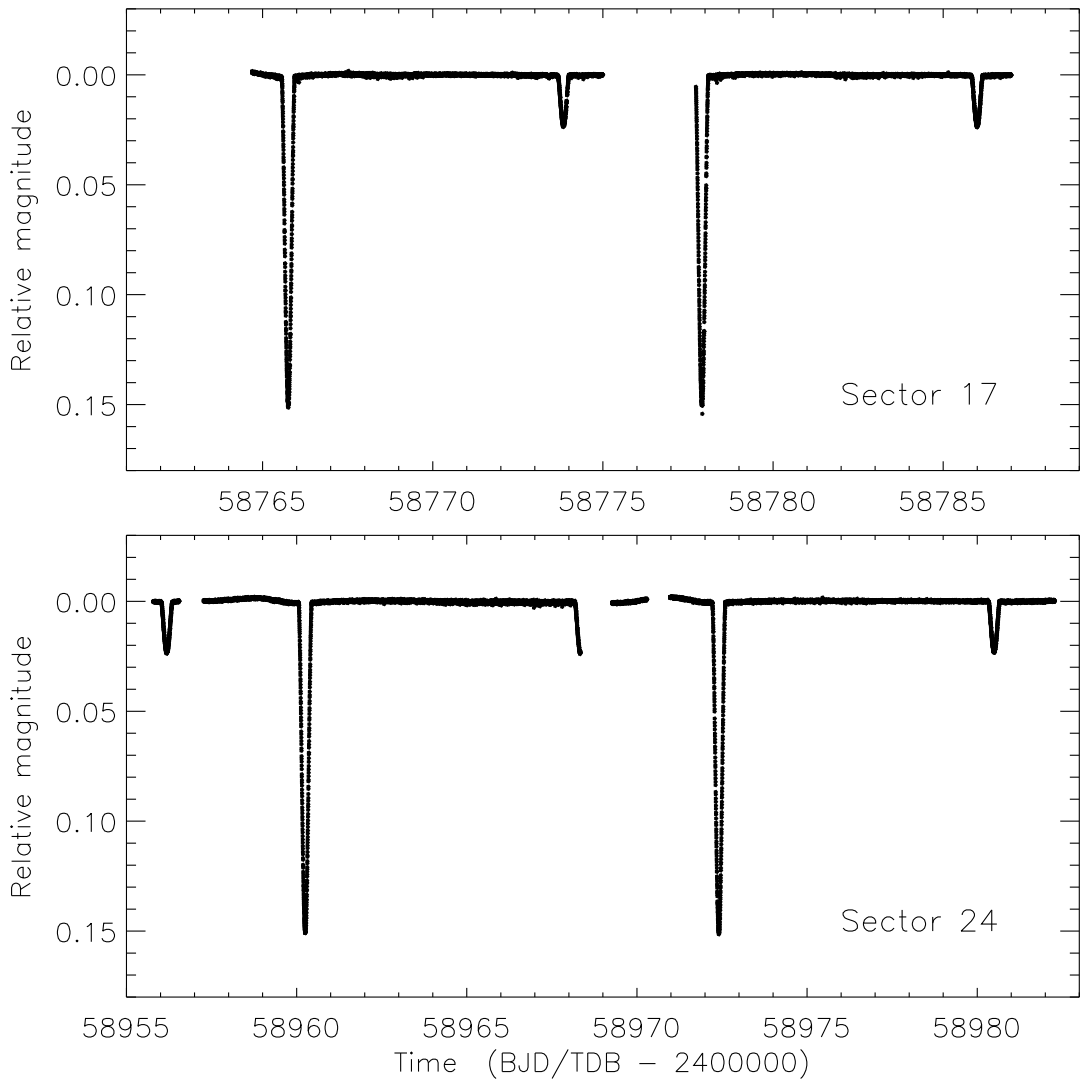} \\
\caption{\label{fig:time} TESS simple aperture photometry of V1022\,Cas. The upper plot
shows observations from Sector 17 and the lower plot shows observations from Sector 24.}
\end{figure}

The main observational material for our analysis comes from the NASA TESS satellite \cite{Ricker+15jatis}, which is discussed in more detail in Paper\,I \cite{Me20obs}. V1022\,Cas was observed twice, with a gap of five months in between, both times at a cadence of 120\,s. Sector 17 was observed between 2019/10/07 and 2019/11/02 and light from V1022\,Cas was incident on camera 2; Sector 24 covered 2020/04/16 to 2020/05/13 and V1022\,Cas was observed using camera 4. There are breaks in the data for download to Earth, and in places where the data are of lower quality.

We used the simple aperture photometry (SAP) light curve \cite{Jenkins+16spie} for both sectors. For Sector 24 we used only the 17\,060 datapoints with no flagged problems (QUALITY $=$ 0). For Sector 17 we relaxed the quality criteria in order to retain the data around the secondary eclipse at JD 2\,458\,774, resulting in a total of 12\,891 datapoints. The light curves from both sectors are shown in Fig.\,\ref{fig:time}.

The majority of datapoints are outside eclipse so contain negligible information on the properties of the system. We therefore trimmed away datapoints more than 1.5 eclipse durations from the midpoint of an eclipse, leaving behind 5444 datapoints in the region of four primary and four secondary eclipses. The errorbars were then scaled so the best fit to the data (see below) has a reduced $\chi^2$ of $\chi^2_\nu \approx 1.0$.


In our analysis we have also made use of the RVs measured by Fekel \etal\ \cite{Fekel++10aj}, which were obtained from 110 high-resolution spectra from three spectrographs. Fekel \etal\ \cite{Fekel++10aj} quoted relative weights for their observations and we have converted these into uncertainties by taking the inverse. This is because inverse-squared uncertainties were too pessimistic for the lower-weighted observations. The errorbars were then scaled to obtain $\chi^2_\nu \approx 1.0$ separately for each star versus the best fit calculated below.


\section*{Light ratio}

Initial analyses of the light curve indicated that the ratio of the radii of the stars was measured to insufficient precision to reach errorbars of 2\% in radius. The ratio of the radii is highly correlated with the light ratio, so an independent measurement of the light ratio can be used to constrain the ratio and thus individual radii (e.g.\ Refs.\ \cite{Torres+00aj} and \cite{Me++07aa}).

Fekel \etal\ \cite{Fekel++10aj} measured a continuum light ratio of $\frac{\ell_{\rm B}}{\ell_{\rm A}} = 0.898$ (where $\ell_{\rm A}$ and $\ell_{\rm B}$ are the wavelength-dependent relative light contributions of the primary and secondary stars, respectively) at 6430\,\AA. However, this measurement does not have an associated errorbar.

Lester \etal\ \cite{Lester+19aj} measured the light ratio of the stars in two ways. First, their interferometric observations give a weighted average of $0.94 \pm 0.04$ in the $K^\prime$ band. Our own calculation from the individual measurements is $0.939 \pm 0.026$, and we have used this value in the following analysis. Second, they found $\frac{\ell_{\rm B}}{\ell_{\rm A}} = 0.95 \pm 0.06$ for the spectral line stengths using the {\sc todcor} algorithm \cite{ZuckerMazeh94apj} on the H$\alpha$ order in their \'echelle spectra.

The interferometric light ratio must be propagated from the $K^\prime$ passband to the TESS passband in order to directly apply it to the TESS light curve. For this we used a code developed to perform this task for contaminating light in transiting planetary systems \cite{Me09mn,Me+10mn,MeEvans16mn,Me+20aa} that accounts for uncertainties in the light ratio and the effective temperature (\Teff) values of the two stars. Using theoretical spectra from the {\sc atlas9} model atmospheres \cite{Kurucz93} we determined a light ratio of $0.981 \pm 0.057$ in the TESS band, and using spectra from the {\sc BT-NextGen} model atmospheres \cite{Allard+01apj} we found $0.984 \pm 0.062$. This consistency is encouraging, and also in good agreement with the light ratio from H$\alpha$. However, the resulting light ratio cannot be regarded as entirely empirical because of the use of theoretical spectra to propagate it between passbands.


\section*{Light curve analysis}

\begin{figure}[t] \centering \includegraphics[width=\textwidth]{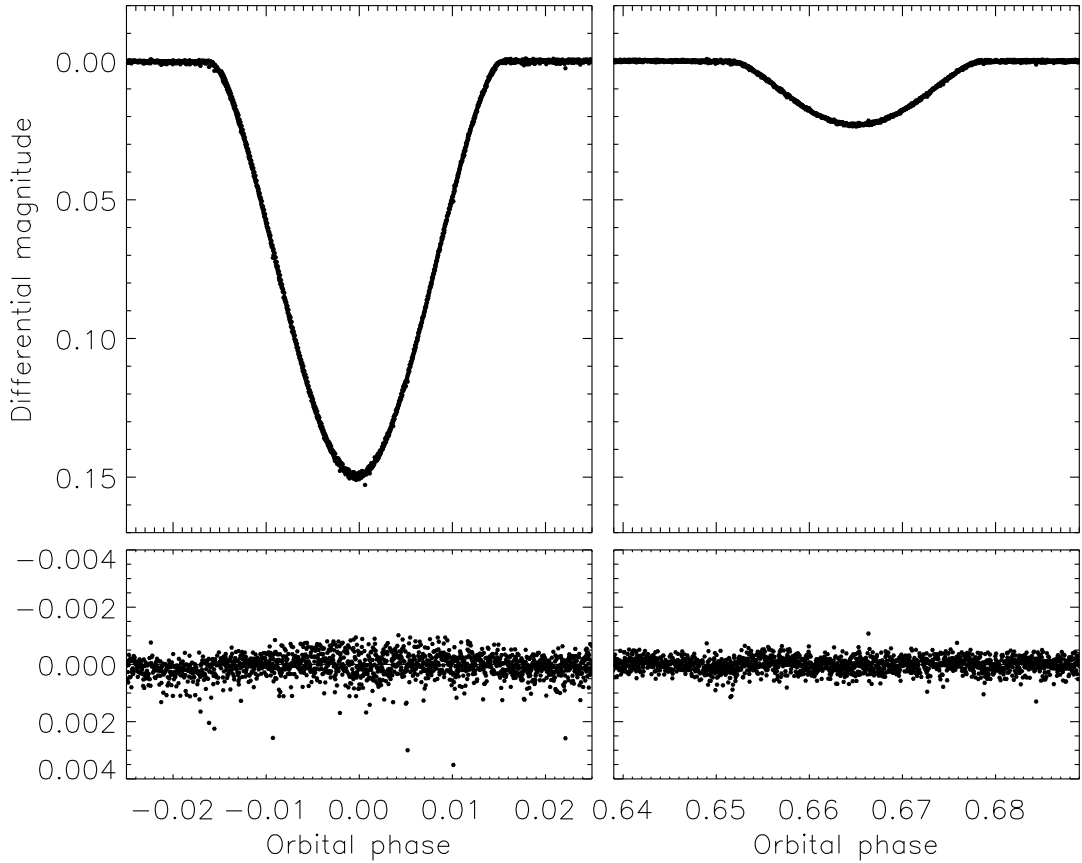} \\
\caption{\label{fig:tess} The TESS light curve of V1022\,Cas around the primary
(left) and secondary (right) eclipses. The {\sc jktebop} best fit is shown using
a solid line, but is obscured by the data. The lower panels show the residuals
of the fit on a larger scale.} \end{figure}

Due to the relatively long orbital period of V1022\,Cas, the stars are approximately spherical so can be modelled using the {\sc jktebop} code\footnote{\texttt{http://www.astro.keele.ac.uk/jkt/codes/jktebop.html}} \cite{Me++04mn2,Me13aa}, for which we used version 41. This code has been shown to agree very well with other codes on well-detached EBs \cite{Maxted+20mn}.

We adopted the standard definition that the primary star is the one eclipsed at the deeper (primary) eclipse. In an EB with a circular orbit, the same area of a star is obscured at both primary and secondary eclipse, so the relative eclipse depths are dictated only by the ratio of the surface brightnesses of the two stars. Hence the primary star is by definition hotter than the secondary star. However, when the orbit is eccentric, the distance between the stars varies and thus the area of star obscured differs between the primary and secondary eclipses. In this case, it is possible for the secondary star to be hotter than the primary. This is the situation for V1022\,Cas, where the primary (star A) is larger and more massive, but the secondary (star B) is hotter.

\begin{figure}[t] \centering \includegraphics[width=\textwidth]{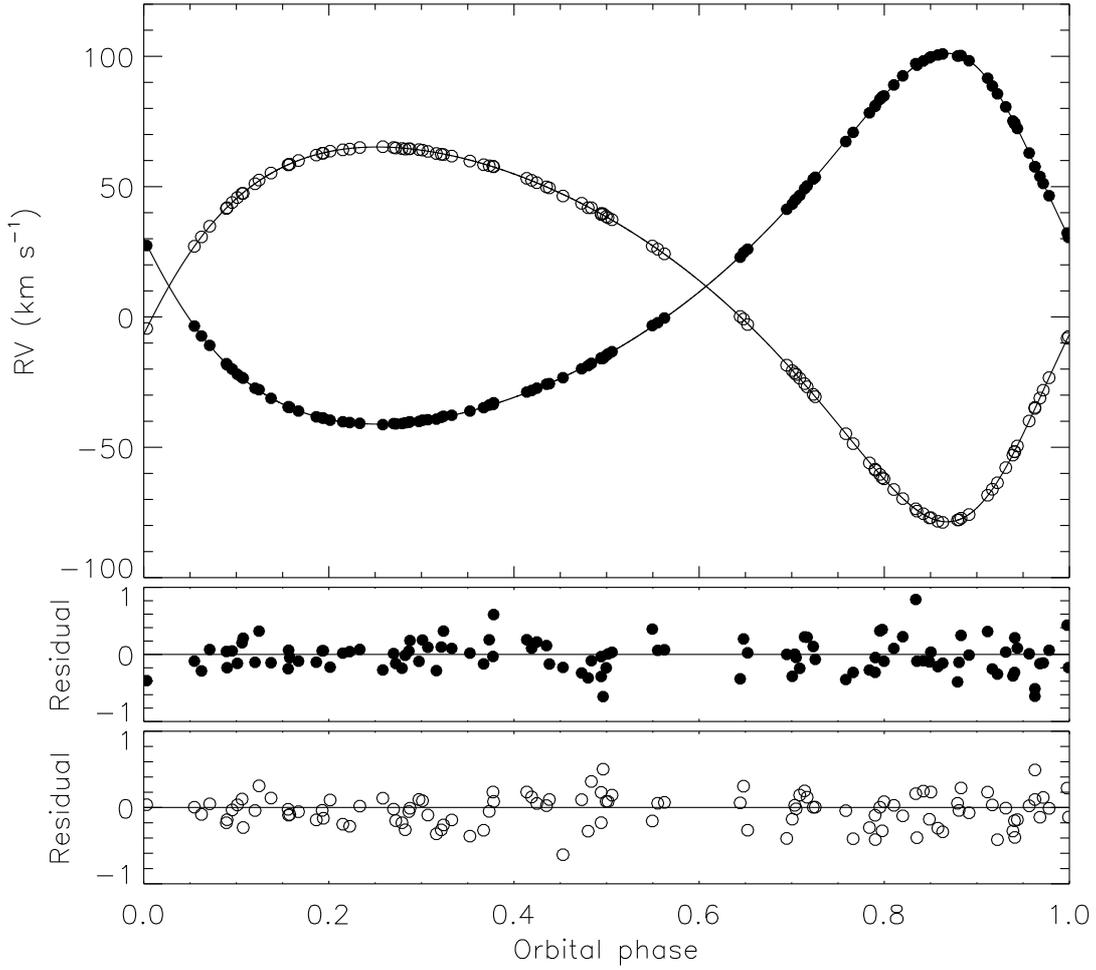} \\
\caption{\label{fig:rv} Spectroscopic orbit of V1022\,Cas using the RVs from
Fekel \etal\ \cite{Fekel++10aj}. RVs of the primary and secondary stars are shown
with filled and open circles, respectively. The best fits from {\sc jktebop} are
shown using solid lines. The lower panels show the residuals of the fit.} \end{figure}

{\sc jktebop} parameterises the fractional radii ($r_{\rm A} = \frac{R_{\rm A}}{a}$ and $r_{\rm B} = \frac{R_{\rm B}}{a}$ where $R_{\rm A}$ and $R_{\rm B}$ are the true radii and $a$ is the semimajor axis of the relative orbit) using their sum ($r_{\rm A}+r_{\rm B}$) and ratio ($k = \frac{r_{\rm B}}{r_{\rm A}}$), and the orbital shape as $e\cos\omega$ and $e\sin\omega$ where $e$ is the eccentricity and $\omega$ is the argument of periastron. These four quantities were included as fitted parameters, as were the orbital inclination, central surface brightness ratio, orbital period, and reference time of primary mid-eclipse. We used the quadratic law for limb darkening, fitted for the linear coefficient, and fixed the quadratic coefficient at a theoretical value \cite{Claret18aa}. Due to the similarity of the stars we required them to have the same limb darkening coefficients.

The TESS data give a very precise ephemeris because the EB was observed in two sectors separated by five months. We fitted the out-of-eclipse brightness of each eclipse using a straight line, in order to avoid systematic errors from any slow variations in brightness remaining in the data. The secondary eclipse occurs at a phase of 0.6641.

\begin{table} \centering
\caption{\em Best {\sc jktebop} fit to the TESS light curve and ground-based RVs of V1022\,Cas.
The 1$\sigma$ uncertainties have been calculated using a Monte Carlo algorithm. The same limb
darkening coefficients were used for both stars. The uncertainties in the systemic velocities
do not include any transformation onto a standard system. \label{tab:lc}}
\begin{tabular}{lr@{\,$\pm$\,}l}
{\em Parameter}                           & \multicolumn{2}{c}{\em Value}    \\[3pt]
{\it Fitted parameters:} \\
Primary eclipse time (BJD/TDB)            & 2458777.91391   & 0.00002         \\
Orbital period (d)                        &      12.1561598 & 0.0000008       \\
Orbital inclination (\degr)               &      82.886     & 0.006           \\
Sum of the fractional radii               &       0.15387   & 0.000055        \\
Ratio of the radii                        &       0.954     & 0.020           \\
Central surface brightness ratio          &       1.0391    & 0.0035          \\
Linear limb darkening coefficient         &       0.2636    & 0.0061          \\
Quadratic limb darkening coefficient      & \multicolumn{2}{c}{0.227 (fixed)} \\
$e\cos\omega$                             &       0.256608  & 0.000014        \\
$e\sin\omega$                             &       0.17660   & 0.00022         \\
Velocity amplitude of star A (\kms)       &      71.112     & 0.024           \\
Velocity amplitude of star B (\kms)       &      71.894     & 0.021           \\
Systemic velocity of star A (\kms)        &      11.775     & 0.019           \\
Systemic velocity of star B (\kms)        &      11.754     & 0.017           \\[3pt]
{\it Derived parameters:} \\
Fractional radius of star A               &       0.07875   & 0.00080         \\
Fractional radius of star B               &       0.07512   & 0.00081         \\
Orbital eccentricity                      &       0.31150   & 0.00011         \\
Argument of periastron (\degr)            &      34.535     & 0.034           \\
Light ratio                               &       0.946     & 0.042           \\
\end{tabular}
\end{table}

In the initial fit, the fractional radii were measured to precisions of only 3.5\% due to the eclipses being partial and the fractional radii being similar. We therefore added both the spectroscopic and the interferometric light ratios discussed above. We made no adjustment from the H$\alpha$ line to the TESS passband because the stars are very similar and the H$\alpha$ line is within the TESS passband. {\sc jktebop} implements this by treating each external light ratio as an observational datapoint in the least-squares fit. The best fit is shown in Fig.\,\ref{fig:tess}. The final results greatly benefit from the inclusion of the external light ratios in the light curve solution.

We then added the RVs from Fekel \etal\ \cite{Fekel++10aj} to the fit. The velocity amplitudes and systemic velocities of each star were fitted separately. The best fit to these data is shown in Fig.\,\ref{fig:rv}. We did not include the RVs measured by Lester \etal\ \cite{Lester+19aj} as they are less numerous and less precise than those from Fekel \etal\ \cite{Fekel++10aj}. The fitted spectroscopic orbit is shown in Fig.\,\ref{fig:rv}.

\begin{figure}[t] \centering \includegraphics[width=\textwidth]{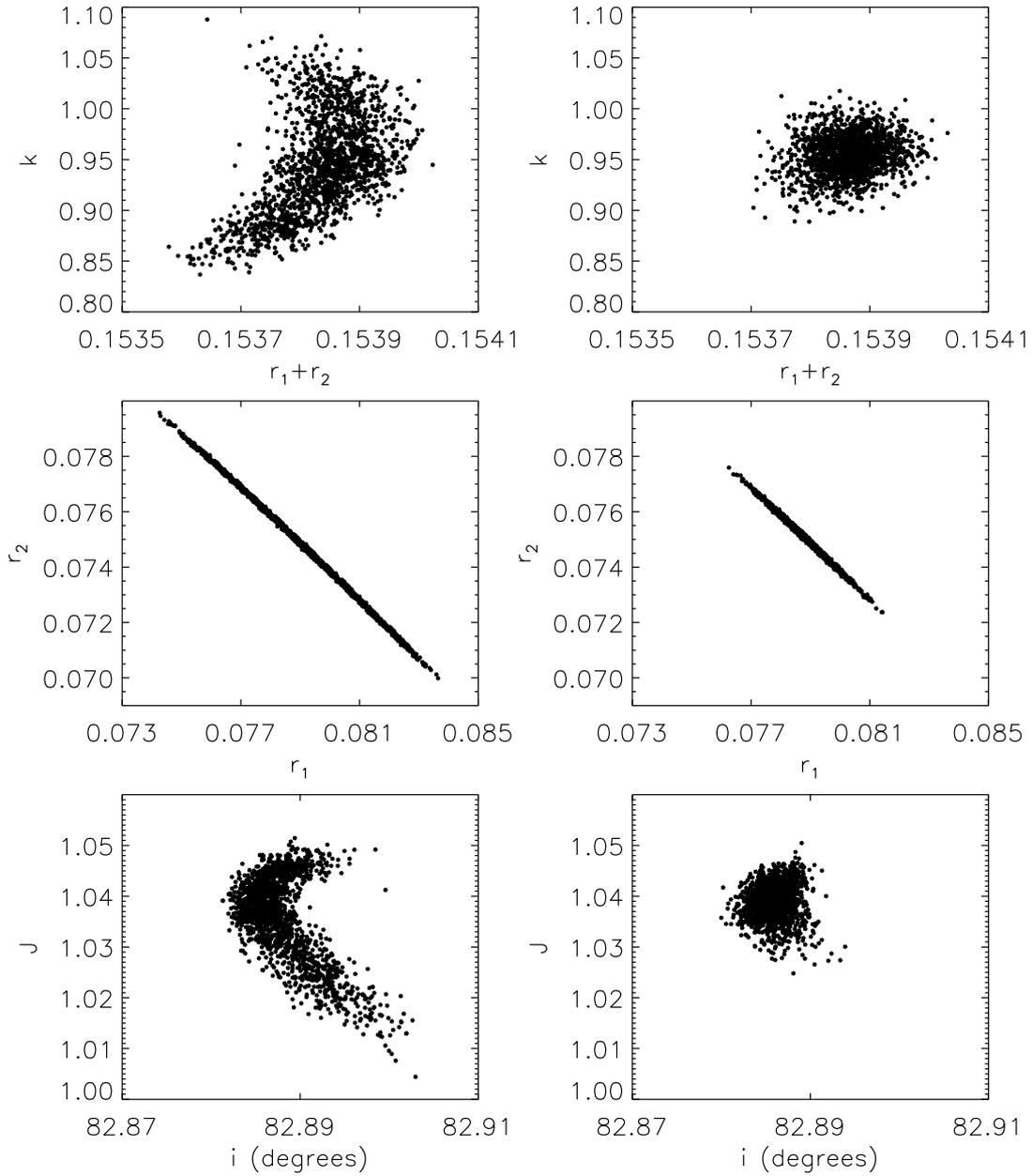} \\
\caption{\label{fig:corr} Plots of the best-fitting values of some light curve
parameters from the Monte Carlo simulations. The left- and right-hand plots
(open circles and filled circles, respectively) show simulations without and
with inclusion of the constraints on the light ratio of the stars.} \end{figure}

The uncertainties of the fit were obtained using the Monte Carlo and residual-permutation algorithms \cite{Me++04mn,Me08mn} implemented in {\sc jktebop}. The Monte Carlo algorithm returned significantly larger uncertainties, so these were used as the final uncertainties. This is because there is no way to account for the uncertainties in the external light ratios in the residual-permutation approach, resulting in underestimated errorbars in this case. The distribution of parameters found in the Monte Carlo analysis is bimodal, with the secondary solution having a larger $r_1+r_2$, $i$, $e\cos\omega$ and $e\sin\omega$. This solution corresponds to a slightly different system configuration but can safely be rejected because it is a much worse fit to the observational data ($\chi^2_\nu \approx 2.0$ versus 1.0 for the main solution). The final parameter values and uncertainties are based on the main solution and are given in Table\,\ref{tab:lc}. The orbital inclination measured from the TESS data is in good agreement with that found by Lester \etal\ \cite{Lester+19aj} from their interferometric observations ($97.1 \pm 0.3^\circ$) once the conversion $i \rightarrow (90^\circ-i)$ is applied to conform to the convention of $0 \leqslant i \leqslant 90^\circ$ for light curve analyses.

For illustration, Fig.\,\ref{fig:corr} shows correlation plots for several combinations of parameters of the light curve fits. The left plots are without, and the right plots are with inclusion of the light ratio constraints. It can be seen that including the light ratios leads to a narrower spread of parameters, and thus more precise measurements of the parameters. It can also be seen that $r_1$ and $r_2$ are very strongly correlated -- this arises because $r_1+r_2$ is much better constrained than $k$ and is the motivation for fitting for the latter parameter combination. The external light ratios help the situation but remain the dominant source of uncertainty in the radii of the stars.


\begin{table} \centering
\caption{\em Physical properties of V1022\,Cas. The \Teff\ values are from Lester \etal\ \cite{Lester+19aj}.
Units superscripted with an `N' are defined by IAU 2015 Resolution B3 \cite{Prsa+16aj}. \label{tab:absdim}}
\begin{tabular}{lr@{\,$\pm$\,}lr@{\,$\pm$\,}l}
{\em Parameter}        & \multicolumn{2}{c}{\em Star A} & \multicolumn{2}{c}{\em Star B} \\[3pt]
Mass ratio                                  & \multicolumn{4}{c}{$0.98912 \pm 0.00044$}  \\
Semimajor axis of relative orbit (\Rsunnom) & \multicolumn{4}{c}{$32.9045 \pm 0.0074$}   \\
Mass (\Msunnom)                             &  1.6263 & 0.0011      &  1.6086 & 0.0012   \\
Radius (\Rsunnom)                           &   2.591 & 0.026       &   2.472 & 0.027    \\
Surface gravity ($\log$[cgs])               &   3.822 & 0.009       &   3.858 & 0.009    \\
Density ($\!$\rhosun)                       &  0.0935 & 0.0028      &  0.1065 & 0.0034   \\
Synchronous rotational velocity (\kms)      &   10.78 & 0.11        &   10.29 & 0.11     \\
Effective temperature (K)                   &    6450 & 120         &    6590 & 110      \\
Luminosity $\log(L/\Lsunnom)$               &   1.020 & 0.034       &   1.016 & 0.030    \\
$M_{\rm bol}$ (mag)                         &    2.19 & 0.08        &    2.20 & 0.08     \\
\end{tabular}
\end{table}

\section*{Physical properties of V1022\,Cas}

Although {\sc jktebop} returned the masses and radii of the components of V1022 Cas corresponding to the best joint fit of the TESS light curve, ground-based RVs, and spectroscopic and interferometric light ratios, we then turned to the {\sc jktabsdim} code \cite{Me++05aa} to calculate these properties as well as other quantities such as luminosity and distance. For this we used the \Teff\ values of the stars given by Lester \etal\ \cite{Lester+19aj}: $T_{\rm eff,A} = 6450 \pm 120$\,K and $T_{\rm eff,B} = 6590 \pm 110$\,K.

{\sc jktabsdim} was used to calculate the full physical properties of the EB using standard formulae, with uncertainties propagated by a perturbation approach. These are given in Table\,\ref{tab:absdim}. They are in good agreement with previous measurements \cite{Fekel++10aj,Lester+19aj} and represent the first determination of the radii of these stars to high precision. The masses are determined to better than 0.1\%, and the radii to approximately 1\% precision. Further improvements in radius measurement could be made by obtaining more precise spectroscopic or interferometric light ratios, which would not be an easy task.

The synchronous rotational velocities for the two stars are 10.8\kms\ and 10.2\kms. The pseudo-synchronous velocities are 21.6\kms\ and 20.4\kms\ (see eq.\,41 from Hut \cite{Hut81aa}). The measured rotational velocities are $v_{\rm A} \sin i = 10.9 \pm 1.2$\kms\ and $v_{\rm B} \sin i = 7.0 \pm 1.3$\kms\ (Ref.\cite{Lester+19aj}), and correcting these to equatorial values gives $11.0 \pm 1.2$\kms\ and $7.1 \pm 1.3$\kms\ (assuming axial alignment). The primary star is rotating synchronously with the orbit, although tidal effects should act to make it rotate pseudosynchronously. Both stars are therefore rotating more slowly than expected from tidal interactions. Together with the significant orbital eccentricity, this means that the system is tidally unevolved.



\section*{Distance to V1022\,Cas}

To determine the distance to the EB we used the $BV$ apparent magnitudes from {\it Tycho} \cite{Hog+00aa}, the $JHK_s$ magnitudes from 2MASS \cite{Cutri+03book}, and the radii and \Teff\ values of the stars. We used two methods \cite{Me++05aa,Me++05iauc}: via the surface brightness calibrations of Kervella \etal\ \cite{Kervella+04aa}, and from theoretical bolometric corrections from Girardi \etal\ \cite{Girardi+02aa}. The optical and near-infrared distances needed a small amount of interstellar reddening to bring them into good agreement -- we found that $E(B-V) = 0.04 \pm 0.04$ (conservative errorbar) is sufficient. The surface-brightness and bolometric-corrections distances agree extremely well. The most precise distance is obtained in the $K_s$ band and is $63.2 \pm 1.1$\,pc from the surface brightness calibration, or $62.7 \pm 1.0$\,pc from theoretical bolometric corrections. The largest contribution to the uncertainties in these values is the $K$-band magnitude, closely followed by the fractional radii and \Teff\ values of the stars. These distances compare well with the value of $63.42 \pm 0.35$\,pc found from the {\it Gaia} DR2 parallax of the system.

A fourth distance estimate is available in this case, using the interferometrically-measured orbital semimajor axis in angular units from Lester \etal\ \cite{Lester+19aj}. Using the small-angle formula and the definition of the parsec, it can be shown that the semimajor axes in angular and physical units are related to the distance according to
$$\left(\frac{a}{\rm au}\right) = \left(\frac{a}{\rm arcsec}\right)\,\left(\frac{d}{\rm pc}\right)$$
The semimajor axis in angular units measured from the interferometry in isolation is $2.390 \pm 0.010$\,mas \cite{Lester+19aj}. When combined with the semimajor axis in Table\,\ref{tab:absdim} the distance is found to be $64.02 \pm 0.27$\,pc. This agrees with the $K_s$-band distance to within 0.7$\sigma$ and with the parallax distance to within 1.4$\sigma$. The four distances are thus consistent and mutually support the measurement principles underlying each.

This good agreement between independent distance determinations is similar to the situation for $\beta$\,Aur \cite{Me++07aa}. TZ\,For has distance determinations from orbital interferometry ($185.9 \pm 1.9$\,pc; Ref.\cite{Gallenne+16aa}) and from \textit{Gaia} DR2 ($183.9 \pm 0.9$\,pc) which also agree to within 1$\sigma$. These three spectroscopic-astrometric-eclipsing binaries provide direct evidence of our good understanding of stellar physics and the local distance scale.



\section*{Comparison with theoretical models}

We compared the measured masses, radii and \Teff\ values to the predictions of several sets of theoretical models to see the level of agreement and to infer the age and metallicity of the EB. The stars are near the end of their main-sequence lifetimes where radius changes quickly, so the age estimates are very precise.

For the PARSEC models \cite{Bressan+12mn} we found a good fit for a solar metallicity and an age of $1930 \pm 40$\,Myr. For the Dartmouth models \cite{Dotter+08apjs} we found the same except for an age younger by 10\,Myr. The Yonsei-Yale models \cite{Demarque+04apjs} prefer an age of 1975\,Myr. In all three cases the solar-metallicity predictions match the observations well and other metallicities do not.

Balachandran \cite{Balachandran90apj} measured a metallicity of ${\rm [Fe/H]} = -0.01 \pm 0.17$ for V1022\,Cas; they were aware that the star is binary but did not account for this in their analysis. This metallicity estimate is roughly solar, but uncertain and probably biased due to neglect of the binarity of the system. It is nevertheless in good agreement with the preference for solar metallicity from the comparison with stellar models. A more discriminating analysis would be possible if higher-precision \Teff\ values and chemical composition measurements were available.


\section*{Summary}

V1022\,Cas has been known to be a double-lined spectroscopic binary for a century \cite{Plaskett+20pdao}. It was discovered to be eclipsing from {\it Hipparcos} photometry \cite{Otero06ibvs}, and an astrometric orbit has recently been published from near-infrared interferometry \cite{Lester+19aj}. It is therefore one of the few spectroscopic-astrometric-eclipsing binaries known.

The TESS satellite has recently observed it twice, giving high-quality light curves of the eclipses. We used these data, together with published RVs, to determine the masses and radii of the stars to high precision for the first time. The masses are measured to 0.1\% using the excellent RVs from Fekel \etal\ \cite{Fekel++10aj}. The radii are determined to 1\%, and the precision of these measurements is limited by the partial nature of the eclipses and the precision of the light ratio measurements from spectroscopy and interferometry.

We have combined these results with published \Teff\ measurements \cite{Lester+19aj} to determine the full physical properties of the system. They are well matched by the predictions of theoretical stellar evolutionary models for a solar metallicity and an age close to 2\,Gyr. A more detailed comparison would benefit from a new spectroscopic analysis yielding photospheric chemical abundances and more precise \Teff\ measurements.

The distance to V1022\,Cas was calculated in four ways: from calibrations of surface brightness versus \Teff\ ($63.2 \pm 1.1$\,pc), from theoretical bolometric corrections ($62.7 \pm 1.0$\,pc), from the interferometrically-measured angular size of the orbit ($64.02 \pm 0.27$\,pc), and from the {\it Gaia} DR2 trigonometric parallax ($63.42 \pm 0.35$\,pc). The agreement between these measurements, obtained in a variety of ways, is very encouraging.


\section*{Acknowledgements}

The following resources were used in the course of this work: the NASA Astrophysics Data System; the SIMBAD database operated at CDS, Strasbourg, France; and the ar$\chi$iv scientific paper preprint service operated by Cornell University.



\end{document}